\documentclass[11pt]{article}
\usepackage[T1]{fontenc}
\usepackage[utf8]{inputenc}
\usepackage[english]{babel}

\usepackage[left=2.5cm,right=2.5cm,top=2.54cm,bottom=2.54cm,includefoot]{geometry}
\usepackage{amsmath,amssymb}
\usepackage{graphicx}
\usepackage{booktabs}
\usepackage{float}
\usepackage{hyperref}
\usepackage{url}

\title{Algorithmic Governance in the United States: A Multi-Level Case Analysis of AI Deployment Across Federal, State, and Municipal Authorities}

\author{%
Maxim Dedyaev\\
National Research University Higher School of Economics, Moscow\\
\nolinkurl{https://orcid.org/0009-0002-2675-5033}\\
E-mail: mdedyaev@hse.ru
}

\date{}
\begin{document}
\maketitle
\vspace{-1.2em}

\begin{abstract}
The rapid expansion of artificial intelligence in public governance has generated strong optimism about faster processes, smarter decisions, and more modern administrative systems. Yet despite this enthusiasm, we still know surprisingly little about how AI actually takes shape inside different layers of government. Especially in federal systems where authority is fragmented across multiple levels. In practice, the same algorithm can serve very different purposes. This study responds to that gap by examining how AI is used across federal, state, and municipal levels in the United States. Drawing on a comparative qualitative analysis of thirty AI implementation cases, and guided by a digital-era governance framework combined with a sociotechnical perspective, the study identifies two broad modes of algorithmic governance: control-oriented systems and support-oriented systems. The findings reveal a clear pattern of functional differentiation across levels of government. At the federal level, AI is most often institutionalized as a tool for high-stakes control: supporting surveillance, enforcement, and regulatory oversight. State governments occupy a more ambiguous middle ground, where AI frequently combines supportive functions with algorithmic gatekeeping, particularly in areas such as welfare administration and public health. Municipal governments, by contrast, tend to deploy AI in more pragmatic and service-oriented ways, using it to streamline everyday operations and improve direct interactions with residents. By foregrounding institutional context, this study advances debates on algorithmic governance by demonstrating that the character, function, and risks of AI in the public sector are fundamentally shaped by the level of governance at which these systems are deployed.
\end{abstract}
\vspace{-0.6em}
\noindent\textbf{Keywords:} artificial intelligence in public governance; AI risk analysis; AI for control; AI for support; algorithmic governance; case study

\section{Introduction}
Contemporary public governance in the United States is undergoing a phase of accelerated algorithmization driven by the active integration of artificial intelligence (AI) technologies into decision-making and administrative processes. In 2025, this trajectory was institutionally consolidated through a series of executive orders issued by the administration of Donald Trump, aimed at deregulating and accelerating the deployment of AI across the public sector (Executive Order on Removing Barriers to American Leadership in Artificial Intelligence, 2025). Within the context of the U.S. federal system, this development creates a distinctive empirical setting for examining how algorithmic technologies become institutionalized in different ways across multiple levels of public authority.

The relevance of this study is further reinforced by the scale and pace of AI diffusion within the U.S. public sector. Available estimates indicate that the use of AI in public administration has been expanding steadily, accompanied by strong expectations of efficiency gains and reductions in administrative costs (SAS Institute, 2024; Deloitte Insights, 2023). Public-sector investment in AI is growing faster than in any other sector, with projected annual growth reaching 19 percent between 2022 and 2027 (IDC Worldwide AI Spending Guide, 2024). The economic potential of algorithmizing administrative processes is equally substantial: according to Deloitte, task automation at the federal level alone could eliminate between 96.7 million and 1.2 billion working hours annually, corresponding to estimated savings of USD 3.3 to 41.1 billion (Deloitte Insights, AI-Augmented Government, 2023).

The scale of AI adoption is particularly pronounced at the federal level. Within a single year, the number of officially registered AI use cases across major U.S. federal agencies more than doubled, including a sharp increase in the deployment of generative models (Federal Artificial Intelligence Use Case Inventory, 2024). This momentum was further reinforced by the adoption of America’s AI Action Plan, which outlines more than ninety AI-related initiatives aimed at modernizing public governance (America’s AI Action Plan, 2025).

At the state level, AI has rapidly evolved from an experimental technology into a core priority of digital governance. By 2025, all U.S. states had either adopted or initiated regulatory and policy measures addressing the integration and governance of AI within public administration (NASCIO, 2024; SAS, 2024).

Similar dynamics are observable at the municipal level, where major U.S. cities have increasingly emerged as hubs of practical AI deployment in routine administrative operations. According to professional associations, a substantial share of large U.S. municipalities already employs AI in at least one functional domain, indicating the progressive institutionalization of algorithmic solutions at the local level of governance (Government Technology, 2024; ICMA, 2024).

Despite the rapid diffusion of AI across federal, state, and municipal levels of public governance in the United States, scholarly understanding of how algorithmic systems transform administrative processes remains fragmented. Existing research tends to focus either on individual levels of authority or on specific technologies and isolated cases, offering limited systematic cross-level comparison of AI deployment practices, their institutional consequences, and the associated risks.

As a result, there is still no empirically grounded understanding of how institutional differences across levels of public authority shape distinct regimes of algorithmic governance within the U.S. model of public administration.

Addressing this research gap, and taking into account the multi-level character of AI deployment in the U.S. public sector, this study advances a comparative analysis of how algorithmic systems transform governance processes across federal, state, and municipal levels of public authority.

The central research question guiding the analysis is as follows: How do institutional differences between federal, state, and municipal levels of authority in the United States give rise to distinct regimes of artificial intelligence use in public governance, and how do these regimes differ in terms of their governance effects and associated risks?

This study makes three key contributions to the fields of public administration and algorithmic governance. First, it offers a cross-level, empirically grounded comparison of how AI is used within different layers of government, moving past studies that focus only on single levels or specific technologies. Second, it proposes a framework that distinguishes between control-oriented and support-oriented AI systems, providing a structured way to understand how institutions vary in the way they use and are affected by algorithmic tools. Third, the study links different governance approaches to varying risk profiles, shedding light on how AI impacts accountability, decision-making power, and discretion across federal, state, and local contexts. Together, these contributions lay the groundwork for future research and help shape current discussions on how to regulate AI in the public sector amid diverse institutional environments.
\section{Theoretical Framework}

\subsection{Digital Transformation of the State and the ``Third Wave'' of the Digital Era}
To conceptualize the observed transformation of public governance, this study draws on the theory of the ``third wave'' of the digital era proposed by Helen Margetts and Patrick Dunleavy. This approach makes it possible to conceptualize the deployment of artificial intelligence not as a continuation of service digitalization, but as a qualitative shift in the architecture of governance that reshapes decision-making processes and the institutional distribution of functions.

Within this framework, the first two waves of digital transformation were associated, respectively, with the digitization of administrative procedures and the subsequent automation of routine bureaucratic tasks. By contrast, the third wave of digital transformation is characterized by the deep integration of AI, big data analytics, and predictive modeling into core governance processes. This integration alters the logic of bureaucratic action and redistributes functional responsibilities among public authorities, algorithmic systems, and private technology providers.

In the logic of the third wave, AI ceases to function merely as an auxiliary tool of automation and instead becomes a structural component of public governance. It directly influences modes of decision-making, configurations of responsibility, and forms of institutional control. This shift is particularly consequential in multi-level systems of public authority, where identical algorithmic technologies may become institutionalized in fundamentally different ways at the federal, state, and municipal levels. The theory of the third wave of the digital era thus provides a conceptual foundation for analyzing algorithmic governance as an institutional process rather than a purely technical transformation. In the present study, this framework is employed to identify how AI is embedded in governance practices across different levels of public authority and to examine the distinct regimes of algorithmic governance that emerge as a result.

This perspective is further developed in the work of Marijn Janssen, who conceptualizes AI adoption as a process of profound organizational and institutional transformation in public governance. From this standpoint, the sustainable use of AI depends less on individual technological solutions than on the state’s capacity to construct coherent data architectures, coordinate organizational change, and adapt governance practices to algorithmic modes of decision-making (Janssen, 2024). Similar conclusions are echoed in systematic reviews of the literature, which emphasize that the effects of AI deployment in public governance are shaped by the interaction of institutional coordination, data quality, and managerial adaptability, rather than by technological sophistication alone (Zuiderwijk et al., 2021).

\subsection{Federalism and AI}
Within the U.S. federal system, the deployment of artificial intelligence follows distinct institutional trajectories. At the federal level, algorithmic systems are primarily embedded through centralized regulatory mechanisms, cross-agency standards, and formal procedures of administrative oversight. As shown by David Freeman Engstrom and his co-authors, this configuration gives rise to a highly coordinated regime of algorithmic governance characterized by standardized rules, centralized authority, and strong inter-agency alignment (Engstrom et al., 2020).

In contrast, at the state and municipal levels, AI deployment is shaped to a much greater extent by conditions of local adaptation, political priorities, and the availability of organizational and financial resources. This results in greater variability of practices, fragmented standards, and substantial differences in both governance effects and risk profiles associated with algorithmic systems. To conceptualize this asymmetry, the study draws on the multi-level governance approach developed by Ignacio Criado, Rodrigo Sandoval-Almazán, and Jorge Gil-Garcia, which conceptualizes AI adoption as a set of interconnected yet institutionally differentiated processes unfolding across multiple levels of public authority (Criado et al., 2025).

Under conditions of federalism, digital transformation thus emerges not as a single, linear trajectory, but as a constellation of parallel institutional pathways. Within this configuration, identical algorithmic technologies contribute to the formation of distinct regimes of algorithmic governance at the federal, state, and municipal levels. This perspective provides the conceptual foundation for a comparative analysis of how differences in institutional coordination, resource endowments, and degrees of managerial autonomy across levels of government are reflected in AI deployment practices, associated risks, and governance outcomes.
\subsection{Algorithmic Accountability, Opacity, and Institutional Risk}
In the United States multi-level federal system where federal, state, and local governments operate with different mandates and resources algorithmic accountability takes on varied forms depending on the level of governance. As Peeters and Schuilenburg suggest, algorithmic systems shift the balance of power by transferring decision-making authority from administrative officials to technical tools that function based on their own logic of data analysis and evaluation.

This shift leads to distinct accountability structures. At the federal level, accountability tends to be organized through centralized oversight bodies and regulations for high-risk AI systems. At the state level, it often relies on a mix of policy-specific protections, auditing mechanisms, and diverse regulatory tools. Meanwhile, local governments facing limited capacity and fragmented oversight, frequently rely on more informal and inconsistent methods of accountability.

One core challenge in algorithmic accountability is the so-called “black box” problem. This concept has both technical and legal dimensions: technically, it refers to the opacity of machine-learning models and their decision-making processes; institutionally, it points to unclear lines of responsibility for outcomes produced by algorithms. Complicating this further is the heavy reliance on private vendors for developing and operating public-sector AI systems. This vendor involvement increases information gaps and limits transparency, independent auditing, and public accountability.

These challenges ranging from algorithmic bias and opacity to weak appeal processes and blurred accountability do not appear evenly across governance levels. Instead, their effects are shaped by each level’s context, capacity, and structure of authority.

These insights suggest that AI use in public governance forms distinct models of algorithmic intervention. Some systems are designed to enhance state control and enforcement, while others support decision-making or improve access to services. How these models function depends on accountability practices, risk management strategies, and the institutional setting in which they are embedded

\subsection{Foundational Concepts for Understanding AI-Driven Governance}
This framework views algorithmic governance as covering a spectrum of activities. The last two areas are particularly associated with artificial intelligence, which acts as a cross-cutting tool that strengthens both analytical power and autonomous capabilities. In contrast, simpler forms of automation don’t necessarily involve AI. This distinction clarifies how general automation differs from governance practices that are significantly shaped by AI.

The analysis also draws on important ideas like algorithmic bias, digital transformation in public services, and institutional fragmentation. Algorithmic bias refers to consistent distortions caused by the way data is collected or models are trained—problems that can mirror or worsen existing social inequalities (Barocas et al., 2019). Institutional fragmentation points to the lack of consistent rules for implementing and overseeing AI, a problem that’s especially noticeable in federal systems.

A complementary view comes from sociotechnical thinking, including the concept of 'AI-enabled transformation' (Tangi et al., 2025). This approach treats AI not just as a stand-alone tool, but as part of a broader network that includes organizations, rules, and governance norms. As a result, the impact of AI depends not just on how it’s designed but also on how institutions adapt decision-making, redefine accountability, and reshape how they interact with the public.

In federal systems, these institutional setups vary across national, state, and local levels, explaining why different approaches to AI emerge in different places.

These concepts highlight that AI’s role in governance can’t be understood just by looking at the technology or individual use cases. The effects depend on how algorithmic systems fit into broader institutional settings, such as governance structures, oversight frameworks, and organizational capacities. This helps explain why the same kind of AI tool might be used for very different purposes in different contexts. The next section introduces a framework for understanding this variation by distinguishing between two main types of AI regimes in public governance: control-focused and support-focused systems.
\section{Analytical Framework: Control- and Support-Oriented AI Regimes}
In this study, the deployment of artificial intelligence in public governance is examined through the lens of two analytically distinct regimes of algorithmic intervention: control-oriented and support-oriented AI regimes. This distinction is derived from the theoretical perspectives discussed above and is employed in a strictly analytical rather than normative sense. Importantly, these regimes are not conceptualized as mutually exclusive technological types, but as analytical categories used to identify the dominant institutional logic through which AI is applied in specific governance contexts.

The control-oriented AI regime is defined as the use of algorithmic systems aimed at strengthening the supervisory and coercive capacities of the state. This regime encompasses cases in which AI is employed to:
\begin{enumerate}
\item detect violations, irregularities, or anomalous patterns;
\item impose sanctions or restrict access to public resources and services;
\item monitor the behavior of individuals or organizations;
\item automate or intensify coercive governance decisions.
\end{enumerate}

By contrast, the support-oriented AI regime includes algorithmic systems whose primary function is to expand the governance and service-delivery capacities of public authorities and, indirectly, of citizens. This regime covers cases in which AI is used to:
\begin{enumerate}
\item expand access to public services;
\item enable predictive prevention of social or economic risks;
\item reduce administrative and transactional burdens;
\item support governance decisions without direct sanctioning consequences for policy recipients.
\end{enumerate}

Empirical cases were classified according to the dominant functional orientation of the algorithmic system, as indicated by the presence of the criteria listed above. In instances where a system combined element of both regimes, priority was assigned to the function that produced the most direct institutional consequences for the subjects of governance intervention.

This classification procedure enables systematic comparison of empirical cases across levels of public authority and facilitates analysis of how the institutional conditions of federalism shape the predominance of control-oriented or support-oriented regimes of algorithmic governance.
\section{Methods}
This study adopts a comparative qualitative research design to examine how artificial intelligence is institutionally embedded in public governance across the United States. Methodologically, the analysis is grounded in qualitative case analysis, which enables the examination of algorithmic systems not as isolated technical artefacts, but as governance instruments operating within concrete organizational, legal, and political contexts.

The choice of a comparative qualitative approach is directly informed by the explanatory nature of the research question, which seeks to account for how institutional differences between levels of public authority give rise to distinct regimes of AI use in government. Addressing this question requires close attention to decision-making logics, organizational adaptations, and institutional effects associated with algorithmic systems dimensions that are difficult to capture through quantitative approaches primarily oriented toward diffusion metrics, performance indicators, or technical characteristics of AI technologies.

Rather than treating individual cases as isolated or idiosyncratic examples, the study proceeds from the assumption that algorithmic governance constitutes an institutional and sociotechnical process. The effects of this process are conditioned by the level of government, the surrounding governance architecture, and prevailing arrangements of accountability. Individual cases are therefore analyzed as empirically grounded manifestations of broader regimes of algorithmic governance. This approach makes it possible to trace how similar classes of algorithmic systems produce divergent governance effects when embedded in different institutional environments, thereby enabling theoretically informed comparison across federal, state, and municipal contexts.

The analytical strategy is guided by an integrated theoretical framework that combines the concept of the ``third wave'' of digital-era governance, a sociotechnical perspective on AI-enabled governmental transformation, and the analytical distinction between control-oriented and support-oriented AI regimes. These concepts are employed explicitly as analytical tools rather than normative benchmarks. They serve to structure the empirical material, guide within-case interpretation, and ensure systematic comparability across different levels of public authority (Fig.~9).

The unit of analysis is a specific instance of AI-based or algorithmic system deployment within a public-sector organization. Each case is conceptualized as an institutionally situated configuration that integrates the functional purpose of the algorithmic system, the organizational context of its use, the form of governance intervention it enables, and the associated risks and observable effects. This configuration-oriented approach allows the analysis to move beyond surface descriptions of technology adoption and to reconstruct the causal pathways through which institutional context shapes governance outcomes.

The study follows a logic of analytic rather than statistical generalization. Its objective is not to provide a comprehensive or representative mapping of all AI applications in the U.S. public sector, but to identify recurring patterns, mechanisms, and institutional configurations through which algorithmic governance is enacted at different levels of public authority. Cross-case comparison is conducted by systematically tracing similarities, contrasts, and contextual variations across cases, thereby supporting theoretically grounded conclusions about differences in AI governance regimes within the multi-level structure of the U.S. state.
\section{Alternative Explanations and Analytical Scope}
A key methodological challenge in comparative research on AI deployment concerns the risk of attributing observed differences in governance outcomes to institutional level alone, while alternative explanatory factors may be at play. This study explicitly addresses three such alternative explanations: resource-based, technology-driven, and domain-specific accounts.

A resource-based explanation would suggest that observed differences in AI governance regimes primarily reflect variations in financial capacity, technical expertise, and administrative scale. From this perspective, control-oriented AI systems at the federal level could be interpreted as a function of superior resources rather than institutional position. While resource disparities are empirically evident, cross-case analysis indicates that resource availability alone does not account for the observed variation. Several resource-rich states and municipalities deploy predominantly support-oriented AI systems, while some comparatively resource-constrained federal agencies operate highly institutionalized control-oriented systems. This suggests that resources condition the scope of implementation but do not determine the dominant governance logic of AI use.

A technology-driven explanation would attribute variation in governance effects to differences in model sophistication, data volume, or computational infrastructure. However, the empirical material shows that similar classes of technologies such as machine learning--based risk scoring, computer vision systems, or generative models are deployed across all levels of government, yet perform markedly different governance functions depending on institutional context. This pattern indicates that technological characteristics are insufficient to explain regime differentiation in the absence of institutional analysis.
A domain-specific explanation would emphasize policy area as the primary driver of AI governance outcomes, suggesting that control-oriented regimes are inherent to domains such as security or taxation, while support-oriented regimes are characteristic of social policy or service delivery. While domain effects are observable, they do not fully account for cross-level variation. Comparable policy domains such as social welfare, healthcare administration, or regulatory enforcement exhibit divergent AI governance regimes across federal, state, and municipal levels, depending on institutional mandates, accountability arrangements, and degrees of administrative discretion.

So, these alternative explanations are analytically relevant but insufficient on their own. The comparative design adopted in this study treats institutional level not as a proxy for resources, technology, or policy domain, but as a causal context that shapes how these factors are combined, constrained, and translated into governance practice. By focusing on institutional embedding rather than isolated variables, the analysis reconstructs the mechanisms through which similar technological capabilities generate distinct regimes of algorithmic governance across levels of public authority. Figure 1 visualizes the analytical and methodological framework of the study, showing how institutional context shapes AI governance regimes through a comparative qualitative design (Fig.~1).\begin{figure}[H]
  \centering  \includegraphics[width=0.9\linewidth]{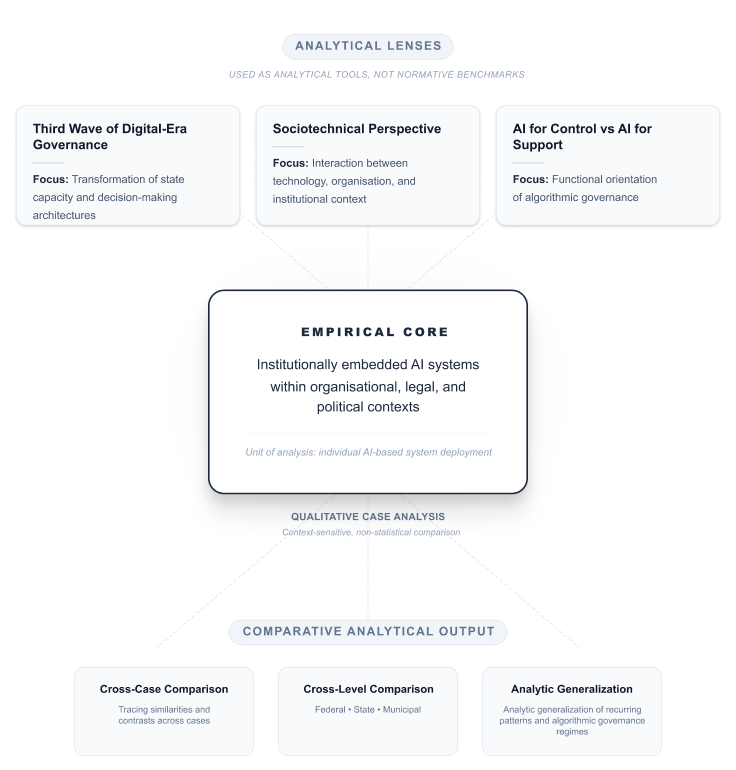}
  \caption{Analytical and Methodological Framework of the Study}
  \label{fig:framework}
\end{figure}
\section{Empirical Analysis: Comparative Patterns of AI-Driven Governance}

\subsection{Federal Level: Scale, Control, and the Institutionalization of Algorithmic Governance}
Artificial intelligence deployment at the federal level of U.S. public administration marks the most expansive and deeply rooted transformation in algorithmic governance across the public sector. Federal agencies benefit from a unique mix of political authority, regulatory power, and resource availability, allowing them to implement large-scale AI systems that go far beyond pilot projects and are firmly embedded in long-standing institutional structures. At this level, AI is used in mission-critical areas such as customs and border control, tax enforcement, national defense, weather forecasting, and space exploration, sectors where AI is helping reshape core aspects of government decision-making.

Analysis of federal use cases shows that AI is primarily viewed as a tool for enhancing security, control, and strategic oversight in high-risk environments. Unlike local and state governments, where AI often addresses specific administrative issues, federal deployments are backed by formal political mandates and are part of national strategies for digital transformation. These strategies include frameworks like the NIST AI Risk Management Framework, which help ensure consistency and long-term stability. At the same time, this approach increases the political and regulatory impact of federal AI systems (Fig. 2).
\begin{figure}[htbp]
  \centering  \includegraphics[width=\linewidth]{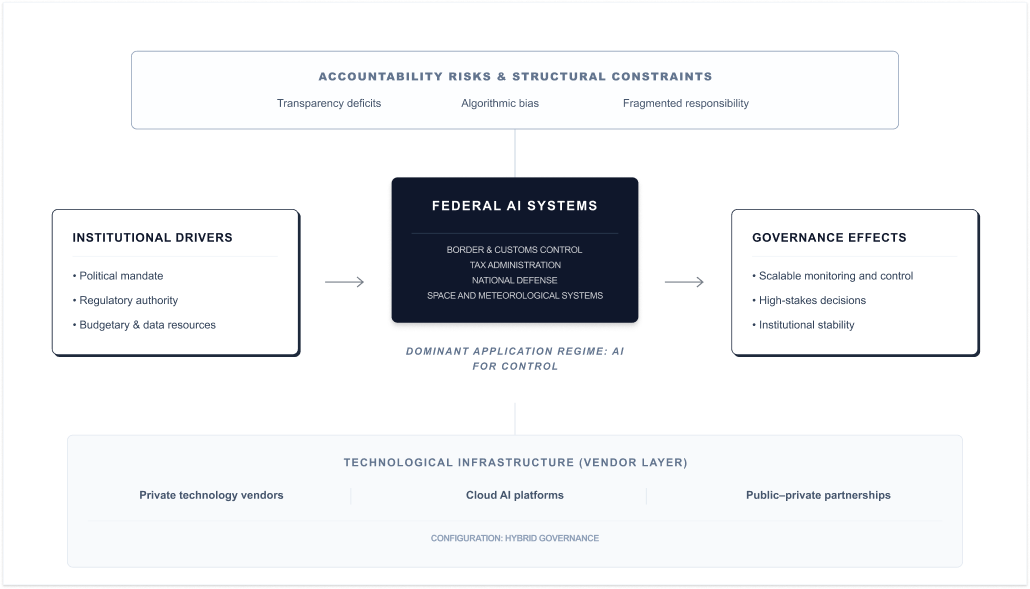}
  \caption{Institutional Drivers, Governance Effects, and Accountability Risks of Federal AI Systems}
  \label{fig:federal-drivers}
\end{figure}

Functionally, most federal AI systems fall under what is known as control-oriented AI. These technologies are used for detecting risks, identifying violations, verifying identities, forecasting threats, and improving oversight. For example, biometric systems are used at the border, predictive models help the IRS detect anomalies, and automated analytics support national defense operations. While this enhances the capacity of government agencies to act quickly and efficiently, it also raises major concerns about transparency, accountability, and algorithmic fairness---key issues in the public discourse on AI governance.

A defining feature of federal AI implementation is its heavy reliance on private-sector technology providers. Case studies show that even high-security agencies depend significantly on commercial cloud services and AI solutions from leading tech firms. This creates a hybrid model of algorithmic governance, where public responsibilities are partly carried out within private digital ecosystems. The result is a more complicated landscape for accountability: independent audits become harder, responsibilities can become fragmented, and it may be difficult to assign blame when algorithmic decisions go wrong.

Together, these federal examples represent the most mature form of algorithmic governance in the United States, reflecting key themes of the third wave of digital-era government. At this level, AI is no longer just a tool to assist administrative tasks, it becomes part of the structure of governance itself. It influences how decisions are made, alters bureaucratic procedures, and shifts how much discretion humans retain in public-sector roles. Nowhere are the contours of an emerging “algorithmic bureaucracy” more visible than at the federal level, where automated systems function as enduring parts of the governance apparatus while remaining formally under human oversight.
Table~1 provides an overview of the federal AI use cases analyzed in this section, including their functional domains, technological characteristics, observed effects, and associated risks (Tab.~1).
\begin{table}[p]
  \centering
  \includegraphics[height=0.84\textheight]{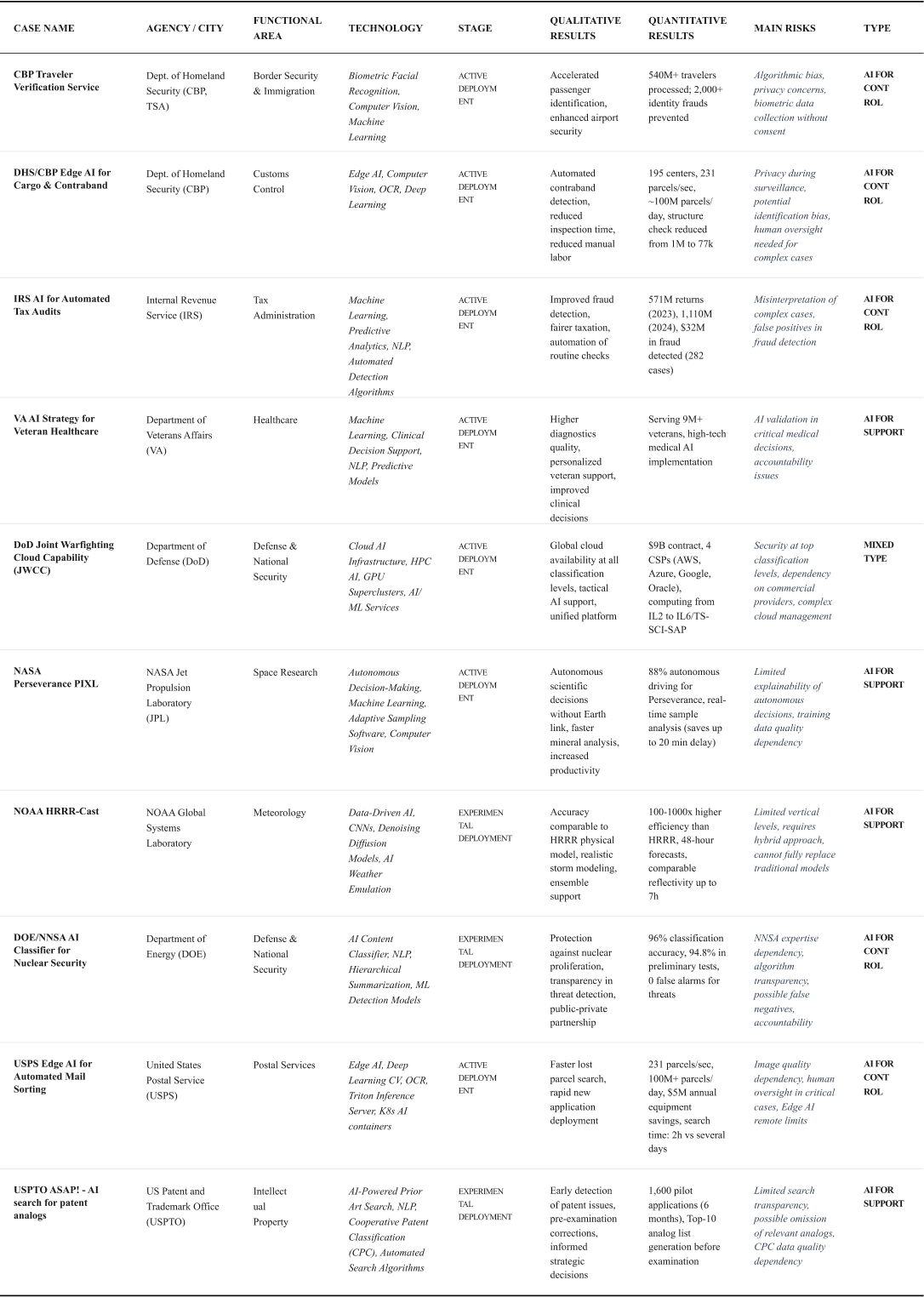}
  \vspace{-0.8em}
  \caption{Summary of Federal AI Use Cases in US Public Administration}
  \label{tab:federal-cases}
\end{table}

\subsubsection{The CBP Traveler Verification Service Case}
The CBP Traveler Verification Service (TVS), deployed by the U.S. Department of Homeland Security and U.S. Customs and Border Protection, represents one of the most institutionally entrenched examples of control-oriented AI deployment at the federal level. The system constitutes a core component of the biometric entry--exit program, whose regulatory foundations have been developed over several decades and were significantly reinforced following the recommendations of the 9/11 Commission. By 2024--2025, TVS operates as a permanent infrastructure of border governance, covering both air and land ports of entry.

Functionally, TVS implements a regime of automated biometric verification. Facial recognition algorithms match live images of travelers against visa and passport databases maintained in government registries. The DHS Final Rule issued in November 2025 formalized mandatory biometric data collection for most categories of non-citizens, thereby consolidating TVS as a fully institutionalized instrument of state control. According to agency reports, the system is applied to hundreds of millions of travelers, enabling identity verification within seconds and substantially reducing operational workload for CBP officers.

From a governance perspective, TVS operates under a logic of default automation, whereby algorithmic verification functions as the primary decision-making mechanism, while human intervention is largely confined to cases of system error or appeal. The case thus illustrates a shift from digital support of border procedures toward autonomous algorithmic surveillance. At the same time, TVS exemplifies the characteristic risks associated with control-oriented AI, including algorithmic bias in biometric identification, privacy concerns stemming from large-scale data collection and retention, and accountability challenges arising from the involvement of private contractors in system development and maintenance.

Analytically, TVS constitutes a paradigmatic case of the federal ``AI for control'' regime, in which an algorithmic system assumes a central role in verification processes and access to state territory. The case demonstrates how a high degree of institutionalization, a strong political mandate, and technological scalability transform AI from a decision-support tool into a structural component of state sovereignty. This transformation, however, simultaneously intensifies challenges related to transparency, accountability, and algorithmic fairness.

\subsubsection{The Joint Warfighting Cloud Capability (JWCC) Case}
The Joint Warfighting Cloud Capability (JWCC), developed by the U.S. Department of Defense (DoD), constitutes a foundational layer of federal infrastructure supporting the transition toward AI-enabled command and control systems. This initiative emerged in response to structural shortcomings in the military’s digital architecture, made particularly evident by the failure of the JEDI program, a single-vendor cloud solution that exposed both political vulnerabilities and technological rigidity inherent in centralized procurement models. In 2022, the DoD adopted a multi-vendor approach, awarding contracts to Amazon Web Services, Microsoft Azure, Google Cloud, and Oracle Cloud. The result was a resilient, distributed cloud system capable of operating across the full spectrum of data classifications, from unclassified to Top Secret.

Unlike domain-specific AI systems, JWCC is not itself an application but an enabling infrastructure. It provides the computational backbone and data environment required to deploy a broad array of AI tools for mission planning, logistics, cyber defense, scenario modeling, and real-time battlefield coordination. It also supports AI operations at the tactical edge, even in low-connectivity settings. By the mid-2020s, many DoD programs involving AI and autonomous systems had integrated JWCC as a system-defining component solidifying its status not as a prototype, but as an institutionalized element of military governance.

From a governance perspective, JWCC embodies a hybrid regime that complicates the conventional distinction between ``AI for support'' and ``AI for control.'' While formally described as a decision-support system governed by principles of human oversight, the platform’s real-time processing capabilities, compressed decision cycles, and integration into operational command chains enable algorithmic outputs to actively shape the space of available choices. This transforms advisory functions into de facto mechanisms of control.

Analytically, JWCC exemplifies the fluid boundary between support and control in high-risk governance contexts. As command-and-control infrastructures become increasingly reliant on commercial cloud-based AI, even ostensibly subordinate algorithmic tools begin to function as structuring agents accelerating decision timelines and constraining discretionary judgment. In this sense, JWCC signals a broader shift from AI as an instrumental aid to AI as a form of governance infrastructure exercising influence not through explicit automation of decision-making, but by shaping the architecture, tempo, and trajectory of military operations.

\subsubsection{The USPTO Artificial Intelligence Search Automated Pilot Program (ASAP!) Case}
The Artificial Intelligence Search Automated Pilot Program (ASAP!), developed by the U.S. Patent and Trademark Office (USPTO), exemplifies a case of algorithmic governance limited by institutional boundaries within the field of intellectual property. The program was launched to relieve growing pressure on the patent review process in highly technical domains, especially those involving artificial intelligence by improving the early stages of application assessment.

ASAP! works by automating the initial search for prior art before formal examination begins. The USPTO’s internal AI system performs semantic analysis of patent databases and compiles a list of relevant prior art for applicants. This changes the context in which applicants decide whether to amend, continue, or withdraw their applications. Although still experimental and used on a limited scale, the algorithm acts as a required epistemic gatekeeper in defining what constitutes novel invention, while officially keeping human examiners in charge of final decisions.

From a governance standpoint, ASAP! sits in a grey area between ``AI for support'' and ``AI for control.'' While presented as a tool that enhances transparency and efficiency, its algorithmic ranking system increases institutional power by narrowing the range of arguments and interpretations available to applicants.
On a deeper level, ASAP! highlights the paradox of what might be called the ``algorithmic democratization of information.'' The system promises earlier and fairer access to critical technical knowledge, yet it also introduces new kinds of informational inequality. This is because the underlying logic of how the AI ranks, filters, and matches documents is not visible to the public. As a result, the traditional adversarial dynamic of patent review is replaced by a process in which both applicants and examiners depend on algorithmically curated layers of information, control over which is concentrated within the institution. In this way, ASAP! represents a subtle, knowledge-based form of algorithmic control. It does not make binding decisions, but it governs what knowledge is accessible and how it is framed.

\subsubsection{Dominant Patterns of Federal-Level AI Deployment}
A comparative review of federal-level AI implementations reveals a consistent and stable functional structure, with several key domains standing out: national security and border control (such as those operated by CBP, DHS, and USPS Edge AI), fiscal and administrative oversight (e.g., IRS and USPTO’s ASAP! system), defense and critical infrastructure (including projects by DoD JWCC, DOE/NNSA, and NASA), and scientific or predictive services (like NOAA’s HRRR-Cast). Although these agencies have diverse missions, they share a common administrative challenge: managing data volumes, speeds, and complexities that outpace the capabilities of traditional bureaucratic systems.

Evidence shows that at the federal level, AI is predominantly used in governance areas where the scale of operations, the urgency of decisions, and acceptable risk levels make exclusive reliance on human judgment impractical. In these cases, AI is not just an auxiliary tool, it becomes an embedded part of core governmental functions. It plays a central role in safeguarding national security, conducting fiscal monitoring, facilitating strategic planning, and improving inter-agency collaboration. This marks a significant difference between how AI is used at the federal level versus its application at state or local levels (Fig. 3).

\begin{figure}[htbp]
  \centering
  \includegraphics[width=\linewidth]{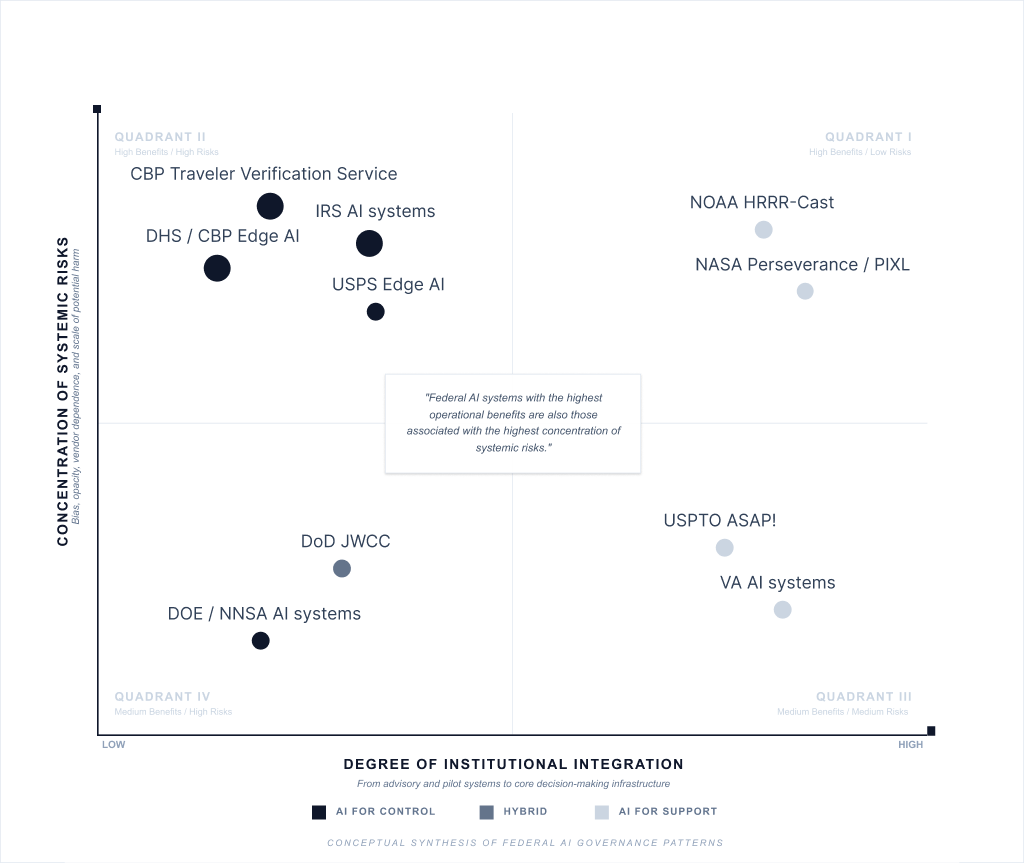}
  \caption{Institutional Integration and Systemic Risk Profiles of Federal AI Systems}
  \label{fig:federal-risk-profiles}
\end{figure}

There are several recurring advantages associated with federal AI use. First, AI allows for massive scalability in administration, enabling agencies to process hundreds of millions of transactions annually without needing equivalent increases in workforce size. Second, AI significantly speeds up vital government functions. Third, these systems enhance the government’s analytical abilities, identifying patterns, risks, and outliers that would be difficult to detect using traditional methods. This is especially apparent in fields like tax compliance, patent processing, and oversight of nuclear and energy systems.

Despite these benefits, federal-level AI deployments also present a cluster of systemic risks. These include algorithmic bias, particularly in facial recognition and identity verification technologies, as well as a lack of transparency due to proprietary systems that remain under partial private sector control. A major structural concern is the federal government’s heavy reliance on commercial cloud providers for running AI systems that handle sensitive or classified data. These dependencies create a new kind of vulnerability where national sovereignty, security, and democratic accountability are increasingly linked to private markets and corporate technology infrastructures.

Overall, the federal government is developing a unique model of AI governance, defined by centralized control, complex infrastructure, and a focus on managing risk at scale. AI is not just used to automate isolated tasks, it is being institutionalized as a core mechanism of governance in areas like national defense, inter-agency intelligence, and the maintenance of critical systems. A consistent trend is the shift from task-specific automation toward continuous feedback loops, where AI systems gather data, generate decisions, and drive administrative action in real time.
From a theoretical standpoint, these findings reinforce key ideas from the third wave of digital-era governance, as well as sociotechnical theories that highlight the transformation of data systems, bureaucratic structures, and the state’s capacity to act. At the federal level, however, these dynamics go one step further. Here, algorithmic governance is not just about making public services more efficient it is also about reinforcing the authoritative and coercive power of the state. Within the framework that distinguishes between ``AI for support'' and ``AI for control,'' federal use cases clearly lean toward the latter. While support-oriented applications do exist, they generally have a more limited and indirect impact on the public.

\subsection{State Level: Variability, Experimentation, and Regulatory Fragmentation}
At the state level, the application of artificial intelligence in U.S. public governance constitutes a domain of high institutional variability, experimentation, and political fragmentation. Unlike federal agencies operating within a relatively coherent strategic framework, state governments develop their own approaches to AI deployment and regulation, shaped by local socio-economic conditions, resource constraints, and political--administrative priorities. As a result, state-level practices of algorithmic governance display substantially greater heterogeneity both in terms of application objectives and institutional consequences.

Empirical analysis indicates that, at the state level, AI is most frequently deployed in the domains of social policy and healthcare. These areas are characterized by particularly strong pressure arising from federal requirements, fiscal constraints, and the dynamics of social problems. Predictive models and AI-based tools are used not only to optimize resource allocation, but also to support the early identification and prevention of social risks. In these cases, AI primarily functions as an instrument of decision support and early intervention, rather than as a mechanism of direct sanctioning.

At the same time, the state level constitutes a key arena of regulatory and institutional innovation. States such as Colorado and Texas have emerged as pioneers in the development of comprehensive AI governance frameworks, incorporating elements of regulatory sandboxes, risk-based oversight, and algorithmic impact assessment procedures. These initiatives give rise to proto-forms of what may be described as ``digital federalism'', within which states experiment with different configurations of accountability, rights protection, and innovation incentives. In contrast to the federal level, where standardized frameworks tend to dominate, state-level approaches are marked by normative competition and the absence of a unified regulatory model. Figure~4 summarizes the institutional drivers, governance effects, and accountability risks that shape AI deployment at the state level, highlighting the hybrid configuration of support- and control-oriented applications (Fig.~4).

\begin{figure}[htbp]
  \centering  \includegraphics[width=\linewidth]{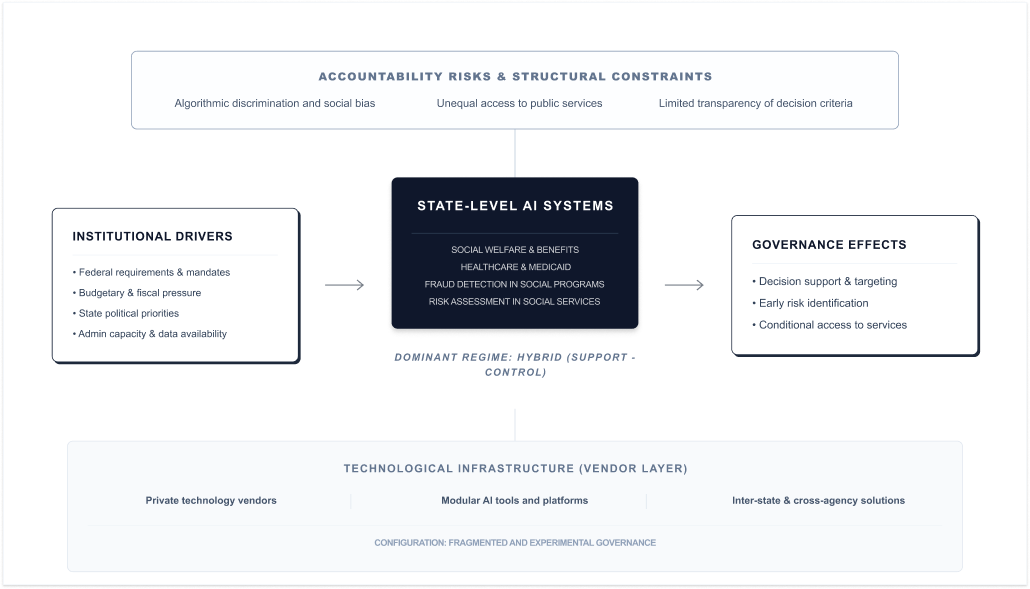}
  \caption{Institutional Drivers, Governance Effects, and Risk Profiles of State-Level AI Systems}
  \label{fig:state-drivers}
\end{figure}

The distinctive character of the state level is particularly evident in its risk profile. Whereas federal AI deployments are primarily associated with risks related to scalability, security, and data sovereignty, and municipal cases tend to raise concerns around privacy and vendor dependence, state-level AI systems are most strongly associated with risks of social inequity and algorithmic discrimination. Across a substantial share of state-level cases, this risk emerges as central: algorithmic models may inadvertently reproduce or exacerbate existing inequalities in access to public services. This is especially salient in sensitive policy areas such as Medicaid administration in Illinois, fraud detection in New York State social programs, and risk assessment systems for youth in state child welfare systems.

Analytically, the state level is best characterized as a hybrid regime of algorithmic governance, combining elements of ``AI for support'' and ``AI for control.'' While most systems are designed to support administrative decision-making and preventive interventions, their progressive institutionalization means that algorithmic assessments increasingly shape citizens’ access to services and resources, thereby acquiring control-like effects. In this sense, the state level occupies an intermediate position between the centralized, control-oriented federal regime and the more pragmatic, service-oriented practices characteristic of municipal governance.
Table~2 provides an overview of the state-level AI use cases analyzed in this section, including their functional domains, stages of institutionalization, observed effects, and associated risks (Tab.~2).
\begin{table}[p]
  \centering  \includegraphics[height=0.84\textheight]{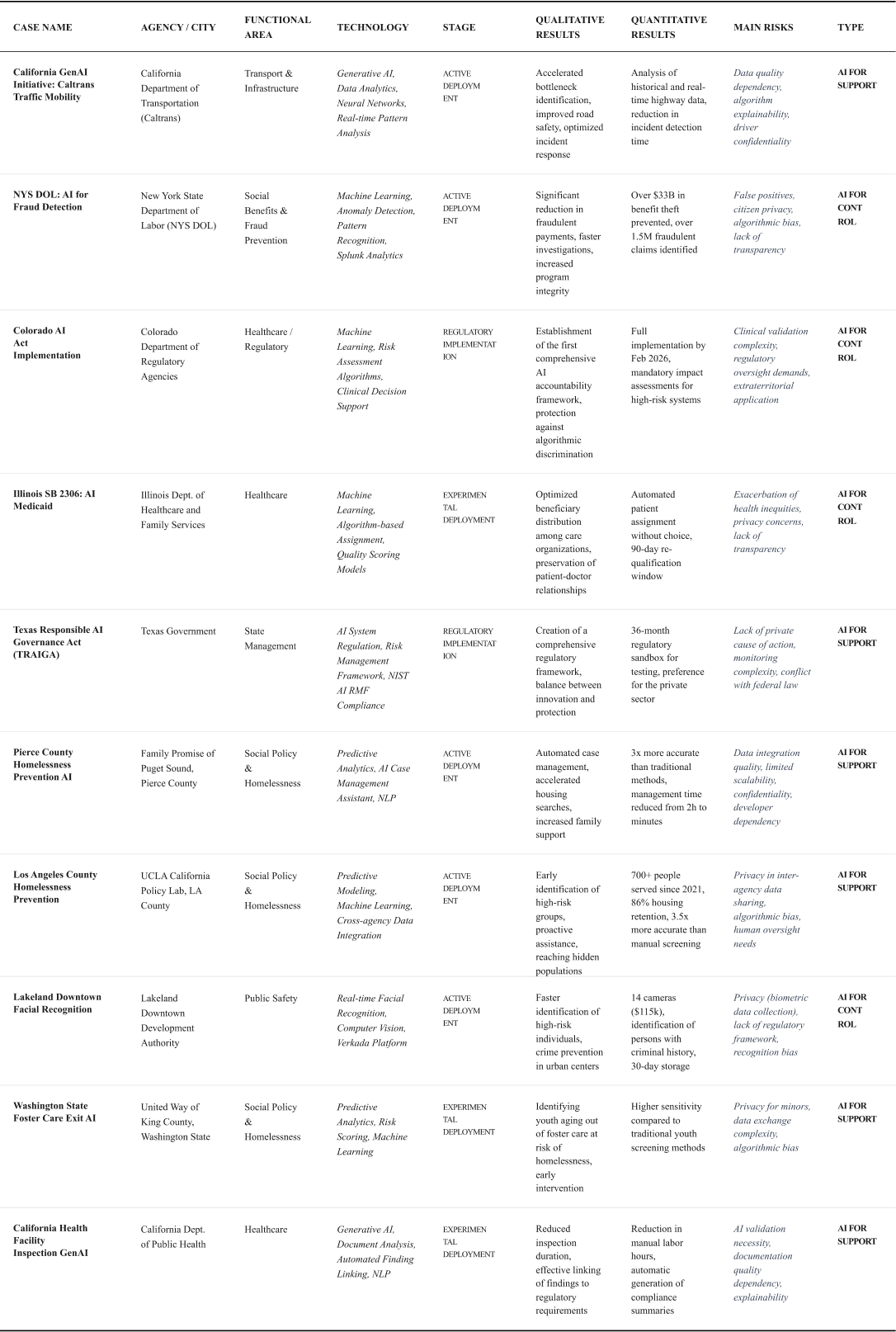}
  \vspace{-0.8em}
  \caption{Summary of State-Level AI Use Cases in US Public Administration}
  \label{tab:state-cases}
\end{table}

\subsubsection{New York State Department of Labor (NYSDOL) Case}
The COVID-19 crisis highlighted deep-seated weaknesses in New York State’s unemployment insurance system. These vulnerabilities came to light during a dramatic spike in claims, widespread data breaches, and an overwhelming number of fraudulent filings. Traditional verification tools were quickly overwhelmed, unable to balance the need for fast processing with fraud prevention. In response, the New York State Department of Labor introduced a comprehensive AI-based risk management system to automate fraud detection and eligibility decisions.

This system combines biometric ID verification, behavioral pattern analysis, and network-based fraud detection. Rather than simply assisting human staff, the algorithms act as the first line of defense: they assign risk scores to applications, suspend or deny payments, and send flagged cases for further review---often without any human input at the early stages. State officials report that this approach helped prevent over \$30 billion in fraudulent claims, demonstrating the system’s fiscal impact.

What began as an emergency measure has since become a permanent fixture in the state’s administrative operations. But as automation becomes embedded in routine workflows, it is also transforming the relationship between citizens and the state, particularly in the sensitive area of welfare services.

Despite its successes, the NYSDOL case reveals significant challenges linked to state-level AI governance. These include the risk of false positives, limited avenues for appeals, unclear decision-making criteria, and technical barriers that disproportionately impact vulnerable groups, such as seniors, people with low incomes, or those lacking reliable internet access. These factors suggest that the push for algorithmic efficiency may increase social vulnerability and reduce equitable access to benefits.

From an analytical standpoint, the NYSDOL case reflects a clear move away from supportive digital tools toward a more controlling AI model in times of crisis. Here, algorithms are not just administrative aids they actively shape who gets access to state benefits. The lack of transparency and unequal access to information heightens the power imbalance between government and citizens, reinforcing critical concerns about how algorithmic systems are reshaping welfare governance.

\subsubsection{Illinois SB 2306 Case}
The implementation of Illinois Senate Bill 2306 (SB 2306) represents a paradigmatic example of hybrid algorithmic governance in state-level healthcare administration. The legislation mandates the use of algorithmic systems to support eligibility assessment, risk stratification, and oversight mechanisms within the state’s Medicaid program, positioning AI as a formally regulated component of welfare governance.

Functionally, the algorithmic framework is designed to optimize administrative efficiency and improve the targeting of medical services by identifying patterns of risk, utilization, and potential abuse. At the same time, the system’s statutory embedding and its direct influence on beneficiary classification endow it with a control function, as algorithmic assessments shape access to healthcare services and the conditions under which benefits are granted or constrained.

Analytically, SB 2306 occupies an intermediate position between ``AI for support'' and ``AI for control.'' While the system is framed as a tool for enhancing service quality and sustainability, its legal codification and operational role in determining eligibility confer substantial institutional authority upon algorithmic outputs. The case thus exemplifies a distinctive state-level configuration in which AI operates as a regulatory mechanism limited in scale compared to federal systems, yet consequential in its effects on citizens’ access to essential public services.

\subsubsection{Los Angeles County Homelessness Prevention AI Case}
The Los Angeles County case constitutes a salient example of support-oriented AI deployment aimed at the preventive management of social crises. In response to the scale and persistence of homelessness, county authorities working in collaboration with the UCLA California Policy Lab developed a predictive system designed to identify households at high risk of housing loss before the onset of crisis conditions.

The algorithmic model employs machine learning techniques and integrates data from multiple administrative sources, including social services, healthcare, housing, and employment programs. Based on predictive risk scores, relevant agencies proactively contact identified households and offer targeted assistance prior to eviction or displacement. Following a pilot phase, the program was scaled up and subjected to a randomized controlled trial (RCT), making it a rare example of systematic empirical validation of algorithmic social interventions in the public sector.

Empirical results indicate a substantial reduction in homelessness among program participants compared to matched high-risk groups not receiving intervention. At the same time, the case highlights risks inherent to support-oriented algorithms, including the invisibility of populations weakly represented in administrative data and privacy concerns arising from cross-agency data integration.
Analytically, this case illustrates a mature configuration of ``AI for support'' augmented by an algorithmic outreach dimension. Unlike control-oriented regimes that restrict access to resources, the predictive model expands the institutional visibility of the state, incorporating previously unengaged populations into the scope of social policy. This allows the case to be interpreted as a shift from algorithmic surveillance toward a form of algorithmically mediated solidarity, in which AI is mobilized for risk prevention and redistribution rather than sanction-based governance.

\subsubsection{Dominant Patterns of AI Deployment at the State Level}
A comparative look at AI applications at the state and county levels reveals a functionally consistent deployment pattern, even amid significant institutional fragmentation. The most prominent use cases are clustered around social policy and welfare programs. For example, predictive tools aimed at preventing homelessness in Pierce County and Los Angeles County, risk-assessment models for youth aging out of foster care in Washington State, and the fraud detection system within New York’s unemployment insurance program all share a central goal: managing social risk and responding in situations of heightened vulnerability.

A second major area of AI use is healthcare and health oversight. This includes Illinois’s algorithmic Medicaid assignment system, Colorado’s risk-based AI regulatory measures, and California’s use of generative AI to inspect health facilities. Unlike federal-level AI, which tends to support strategic control, these state applications mainly focus on easing administrative burdens, improving service consistency, and addressing inefficiencies in health-related processes.

A third category consists of overarching regulatory initiatives like the TRAIGA framework in Texas and Colorado’s AI Act. These programs set broader governance rules for algorithmic systems and portray states as testing grounds for different models of accountability, innovation, and rights protection within a regulatory patchwork.

Across different policy sectors, state-level AI implementations tend to offer recurring benefits. In welfare programs, AI can sharpen the focus on high-risk populations and improve resource targeting. In healthcare and infrastructure, it accelerates problem detection and enhances inter-agency coordination. Regulatory efforts, meanwhile, are building expectations around AI oversight, such as auditability, risk controls, and maintaining human supervision.

Yet, the state level is also where algorithmic inequality and systemic fragmentation are most pronounced. Cases in New York and Illinois reveal how models built on historical trends and performance metrics can reinforce existing geographic and social disparities. These problems are worsened when diverse administrative data are integrated, making data errors or breaches more likely to compound vulnerabilities.

What sets the state level apart is a unique pattern in how AI is used to govern social resource distribution. While federal AI is tied to sovereign and security functions, and municipal AI often supports operational tasks, states and large counties are emerging as rule-makers for algorithmic access to housing, healthcare, and welfare services. A common feature is AI's role in fine-tuning how the welfare state operates---through classifying vulnerability, adjusting intervention levels, and navigating the balance between fiscal constraints and social justice expectations (Fig.~5).

\begin{figure}[htbp]
  \centering
  \includegraphics[width=\linewidth]{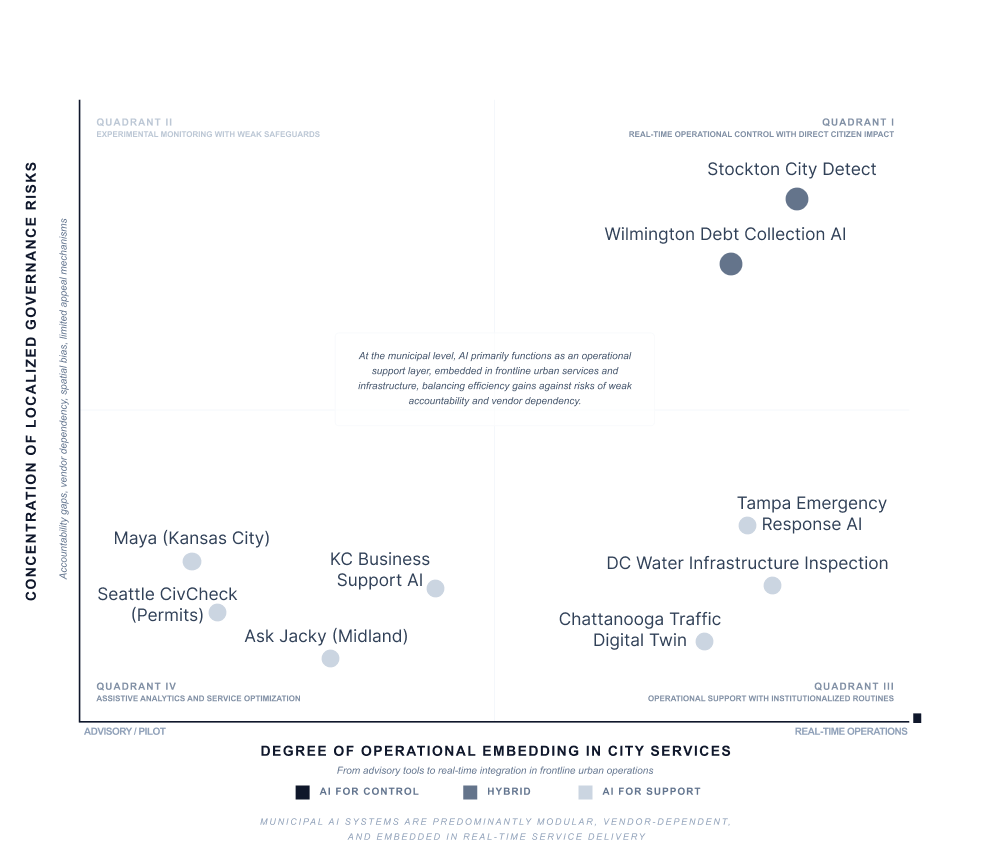}
  \caption{Distributive Integration and Social Risk Profiles of State-Level AI Systems}
  \label{fig:state-risk-profiles}
\end{figure}
From an analytical perspective, the state level is best characterized by a hybrid configuration combining elements of ``AI for support'' and ``AI for control.'' Systems such as those deployed by NYSDOL, Medicaid administration in Illinois, and parts of the Colorado regulatory framework gravitate toward control-oriented logics, while homelessness prevention initiatives and youth support programs exemplify support-oriented regimes. Crucially, the boundary between these regimes remains fluid: identical technical approaches may operate as instruments of sanction or support depending on institutional design, appeal procedures, and political--administrative priorities rather than on model properties alone.

\subsection{Municipal Level: Service-Oriented Algorithmization and Local Experimentation}
The deployment of artificial intelligence at the municipal level of U.S. public governance unfolds under conditions of constrained resources and a high degree of institutional experimentation. Paradoxically, it is precisely this least resource-endowed level of government that has become a central testing ground for the practical algorithmization of urban services. Large cities and mid-sized municipalities increasingly integrate AI into everyday administrative routines ranging from permitting and citizen requests to infrastructure management, transportation systems, and emergency services.

Industry reports by ICMA and Government Technology indicate that a majority of U.S. cities with populations exceeding 100,000 residents already employ at least one AI-based solution, while the total number of pilot and operational projects by 2024--2025 reaches into the hundreds. In contrast to the federal and state levels where AI is embedded in security, regulatory, and welfare governance---municipal AI operates as part of the operational fabric of the city, simultaneously affecting multiple heterogeneous yet routine processes.

The functional profile of the ten municipal cases analyzed in this study demonstrates pronounced heterogeneity. The sample includes systems designed to accelerate permitting and planning procedures, AI-assisted emergency response tools, algorithmic assistants for firefighters and paramedics, multilingual civic chatbots, computer vision applications for infrastructure inspection, digital twins for traffic management, as well as selective uses in law enforcement and revenue administration. This diversity reflects not strategic unification, but rather the adaptation of AI to specific administrative bottlenecks characteristic of urban governance.

Applying the analytical distinction between ``AI for support'' and ``AI for control'' reveals a strong orientation toward support at the municipal level. Eight of the ten cases primarily aim to enhance frontline capacity and improve service responsiveness for residents. Only a limited subset, such as Stockton City Detect and Wilmington Debt Collection can be classified as control-oriented in a strict sense, where algorithms are used for violation detection or targeted enforcement. This distribution stands in sharp contrast to the federal level, dominated by control-oriented AI regimes, and to the state level, where the balance between support and control remains institutionally contested.

The institutional architecture of municipal AI governance is shaped above all by resource fragmentation and vendor dependence. None of the cities examined maintains a fully internal development cycle for complex AI systems. Solutions are typically procured as off-the-shelf products from commercial vendors, implemented through partnerships with universities, nonprofits, or philanthropic initiatives, or deployed in hybrid configurations combining cloud-based services with limited local customization. While this lowers entry barriers for municipalities, it simultaneously generates persistent technological dependencies, whereby the evolution of urban services is partially shaped by vendor product logics rather than solely by municipal policy priorities.

A further defining feature of the municipal level is the radically localized nature of data. Unlike federal agencies relying on national registries or states drawing on large administrative datasets, municipal AI systems operate on data generated directly within the urban environment: images of buildings and public spaces, video feeds from inspection robots in sewer systems, traffic telemetry, citizen service logs, and call-center recordings. The quality and completeness of these local datasets directly condition algorithmic performance, rendering data governance a critical, yet often underdeveloped municipal capability.

Municipal AI governance is also characterized by rapid iteration cycles and a relatively high tolerance for managerial risk. Nearly all cases examined originated as pilot projects with limited scope and short testing horizons. This approach allows cities to quickly validate hypotheses, scale successful applications, and discontinue ineffective initiatives. However, accelerated deployment is frequently accompanied by weak institutionalization of appeal and audit mechanisms. When algorithmic errors occur particularly in enforcement or billing contexts the consequences are felt immediately by residents, while formalized procedures for contestation and independent review are often absent or remain underdeveloped.

Finally, the municipal level exhibits a distinctive configuration of public visibility and accountability. On the one hand, city-level AI projects are relatively visible in media and professional networks, with municipal administrations actively showcasing successes, participating in GovTech competitions, and publishing headline performance metrics. On the other hand, this visibility is rarely matched by robust regulatory infrastructure. Unlike some states, municipalities generally lack mandatory algorithm registries, standardized algorithmic impact assessments, or dedicated complaint mechanisms for AI-driven decisions. The resulting pattern is one of service-oriented algorithmic governance, in which AI functions primarily as a tool for operational support and service modernization, while operating under conditions of fragmented and ``soft'' transparency (Fig.~6).

\begin{figure}[htbp]
  \centering  \includegraphics[width=\linewidth]{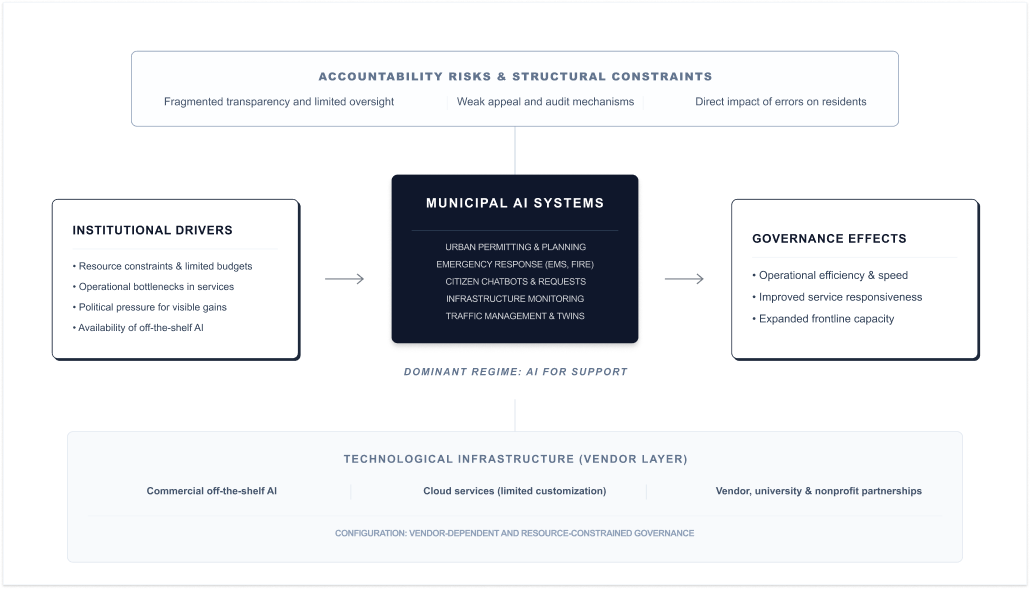}
  \caption{Institutional Drivers, Governance Effects, and Accountability Constraints of Municipal AI Systems}
  \label{fig:municipal-drivers}
\end{figure}Table~3 provides an overview of the municipal AI use cases analyzed in this section, including their functional areas, implementation stages, observed effects, and associated risks (Tab.~3).
\begin{table}[p]
  \centering  \includegraphics[height=0.88\textheight]{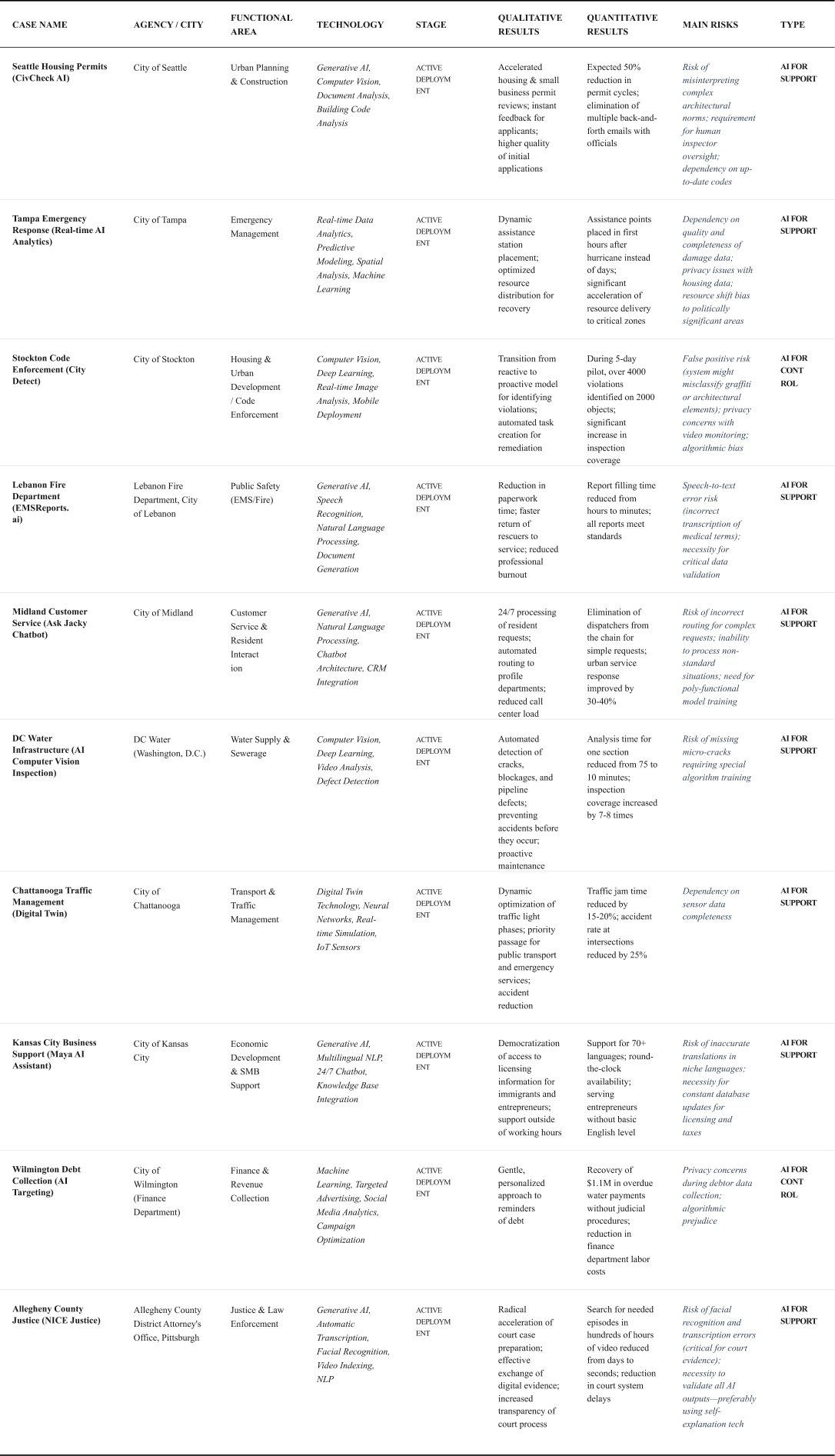}
  \vspace{-0.9em}
  \caption{Summary of Municipal AI Use Cases in US Public Administration}
  \label{tab:municipal-cases}
\end{table}

\subsubsection{Seattle Housing Permits (CivCheck AI) Case}
The implementation of CivCheck AI in Seattle's housing permit process exemplifies a standard support-oriented AI use at the city level. The system was introduced to tackle long-standing problems such as administrative delays, extended application review periods, and frequent errors in architectural submissions---all of which limited the city’s permitting capacity during a housing crisis. CivCheck AI was embedded in the city’s permitting platform as part of the Permitting and Customer Trust initiative, serving as a preliminary review assistant. Technically, it acts as an AI copilot by using computer vision and document analysis to evaluate architectural plans for completeness and compliance with local codes, providing structured feedback before a human review takes place. According to the city, this tool could cut down processing time and reduce administrative workload without changing the official role of inspectors.

From a governance standpoint, the case reflects common challenges in municipal AI implementation: reliance on third-party vendors, the need for constant updates in response to policy changes, and uncertainties around how compliance costs are shared between the city and applicants. These challenges show that even support-focused AI tools require ongoing institutional and budgetary commitment.

From an analytical perspective, CivCheck AI clearly fits the ``AI for support'' model. It doesn’t make binding decisions but restructures the process by adding a two-step review in which the AI aids rather than replaces human judgment. Institutionally, this case illustrates how municipal AI is used to increase efficiency and predictability, rather than shift authority or intensify control.

\subsubsection{Lebanon Fire Department (EMSReports.ai) Case}
The EMSReports.ai tool used by the Lebanon Fire Department is another example of a support-focused AI system, designed to improve administrative performance in emergency services. Launched in 2024 and formally recorded in the city’s public AI registry, the system automates emergency medical services documentation by combining speech recognition with generative AI. Emergency personnel speak incident details aloud, and the system converts the speech into structured reports, checks them for completeness, and uploads them to existing systems. Data from the city suggests this has cut report preparation time by about 75\%, freeing up resources for emergency response and training.

Yet, this case also reveals typical risks of municipal AI in sensitive settings, such as transcription mistakes, excessive reliance on AI-generated reports, and declining documentation skills among staff. Although the system is built to meet HIPAA standards, its effectiveness hinges on strict human oversight and verification processes.

Analytically, EMSReports.ai also fits into the ``AI for support'' category. It doesn’t guide medical or dispatch decisions but lessens the paperwork burden. Its listing in the public AI registry makes it a noteworthy instance of how cities can try to compensate for limited AI accountability through transparency measures and formal oversight.

\subsubsection{Chattanooga Traffic Management (Digital Twin) Case}
The Chattanooga traffic management project shows how AI-driven digital twins can be used to manage municipal infrastructure in real time. With support from a strong local digital infrastructure and academic partners, the city has built a virtual model of its road network to monitor, simulate, and optimize traffic flows. This digital twin uses sensor networks, edge computing, and AI tools to analyze traffic at the level of individual streets and intersections. The system’s algorithms adjust traffic lights, prioritize emergency and public transit, and flag potential safety issues. Early tests suggest that this AI-driven traffic system can cut energy use and emissions while maintaining or improving traffic flow.

However, the project also demonstrates the structural risks of using AI to manage infrastructure. Its success depends on the availability of sensors and stable data connections, and localized optimization might worsen traffic in outlying areas. Also, decisions about what to optimize, such as speed, safety, or environmental goals are political, not purely technical.
Analytically, Chattanooga’s system lies somewhere between ``AI for support'' and ``AI for control.'' While it exerts real-time algorithmic control over city infrastructure, its purpose is framed as enhancing safety and promoting sustainability. Overall, the case shows how AI systems managing city infrastructure become tools of urban policy, where the distribution of benefits and risks reflects governance decisions rather than neutral tech performance.

\subsubsection{Dominant Patterns of AI Deployment at the Municipal Level}
Comparative analysis of municipal AI cases demonstrates that cities constitute the most heterogeneous and experimental layer of algorithmic transformation in public service delivery. Unlike federal and state authorities, municipalities rarely pursue large-scale or integrated AI governance regimes. Instead, algorithmic systems are introduced in a targeted manner, addressing specific, routine urban processes rather than reshaping overarching institutional architectures.

The dominant pattern at the municipal level is the deployment of AI at the frontline of interaction between government, residents, and urban infrastructure. This is evident in the acceleration of permitting procedures (Seattle CivCheck), automation of emergency service documentation (EMSReports.ai in Lebanon), algorithmic management of transport and engineering infrastructure (Chattanooga traffic digital twins, DC Water inspection systems), expansion of communication channels with residents and businesses (Ask Jacky in Midland, Maya in Kansas City), as well as in selected cases of localized enforcement and fiscal control (Stockton City Detect, Wilmington debt collection). Across these applications, AI is embedded directly into operational workflows and used to increase speed, predictability, and administrative responsiveness (Fig.~7).

\begin{figure}[htbp]
  \centering
  \includegraphics[width=\linewidth]{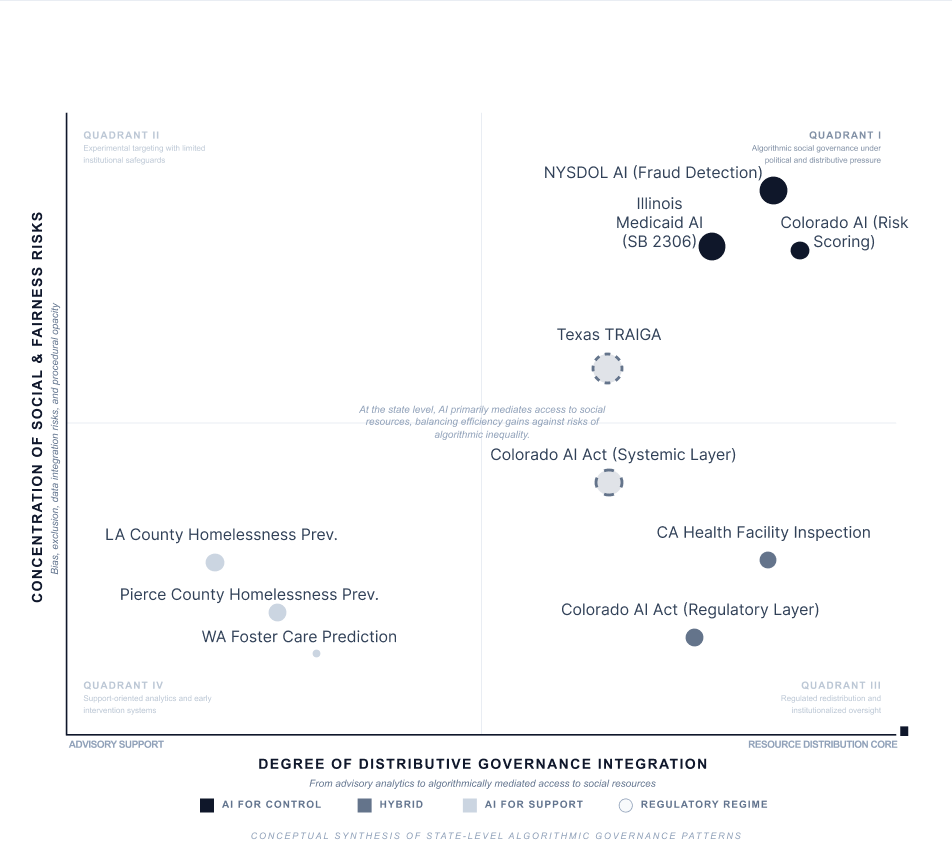}
  \caption{Distributive Integration and Social Risk Profiles of Municipal-Level AI Systems}
  \label{fig:municipal-risk-profiles}
\end{figure}

A unifying characteristic of municipal AI deployments is their proactive orientation. In contrast to traditional reactive models of city governance, algorithmic systems enable anticipatory intervention by identifying issues before they escalate into formal complaints or system failures. In Stockton, code violations are detected prior to citizen reports, while in DC Water infrastructure defects are identified well ahead of accidents. This produces an additional digital layer of urban management, redistributing operational burden from human staff to algorithms and compressing decision-making cycles. This dynamic can be described as a form of real-time micro-management of the city.

Municipal AI systems are predominantly modular and task-specific, in contrast to the platform-based architectures more typical of state-level governance. This configuration reflects not only fiscal and staffing constraints, but also the pragmatic logic of urban administration, which prioritizes rapid problem-solving over long-term system integration. Cities function as experimental environments characterized by small-scale deployments, short pilot cycles, and selective scaling of successful solutions. At the same time, this results in a fragmented sociotechnical ecosystem, where individual AI tools operate in relative isolation from one another.

Municipal governments across different cases report significant drops in transaction costs. Tasks like processing reports, applications, and inspections are now completed much faster, infrastructure is used more efficiently, and access to services has improved, especially for multilingual and underserved communities. AI systems also help make administrative decisions more consistent by standardizing how information is handled and minimizing differences caused by human judgment. In this way, municipal AI helps produce practical governance insights that would otherwise be expensive or difficult to achieve through traditional reforms.

However, municipal use of AI also comes with a unique set of risks. One of the biggest challenges is the lack of strong institutional systems for holding these technologies accountable. In many cities, audit processes, registries for high-risk systems, and formal appeal mechanisms are either missing or poorly developed. Because of this, many municipal AI systems function in a regulatory grey area. The heavy reliance on private tech vendors adds another layer of vulnerability, placing much of the control outside city hands and making it harder to evaluate algorithmic fairness or performance.

Another concern is the potential for these systems to deepen social and geographic inequalities. For example, automated enforcement tools may disproportionately impact low-income neighborhoods, traffic AI may divert congestion to less connected areas, and budget-related AI systems may put added pressure on already vulnerable residents. These outcomes aren’t usually part of the system’s intended goals, but they often arise as unintended results of optimization processes.

Compared to the federal and state levels, municipal governments stand out. While federal agencies focus on enforcement and risk management, and states tend to build formal rules for distributing benefits, cities work at ground level. They deal with the day-to-day issues people experience like traffic, permits, broken infrastructure, and access to services. Because of this, city-level AI use is largely geared toward helping, not controlling. The technology earns public trust by providing clear and immediate improvements to urban life.
In theory, AI in cities is less about replacing human decision-making and more about building the underlying digital systems that support government work, such as automation and data analysis. These tools rarely make final decisions on their own but significantly reshape how decisions are prepared. This shifts influence to earlier stages, like how models are designed and how data is managed. In this environment, the success and trustworthiness of city-level AI depend less on how advanced the technology is and more on the institutional and organizational context it’s used in.

\section{Conclusion: A Three-Level Structure of Algorithmic Governance in the United States}
This study systematically examined 30 AI deployment cases across three levels of public authority in the United States---federal, state, and municipal---using a comparative qualitative case analysis. The results demonstrate that algorithmic governance in the U.S. does not constitute a unified or homogeneous regime. Instead, AI adoption follows a level-dependent institutional configuration, in which the same classes of AI technologies perform qualitatively different governance functions depending on their embedding within federal, state, or municipal administrative architectures.

Figure~8 synthesizes the core empirical findings by conceptualizing U.S. algorithmic governance as a three-level governance stack, characterized by distinct functional orientations, institutional roles, and dominant AI regimes.
\begin{figure}[p]
  \centering  \includegraphics[height=0.74\textheight]{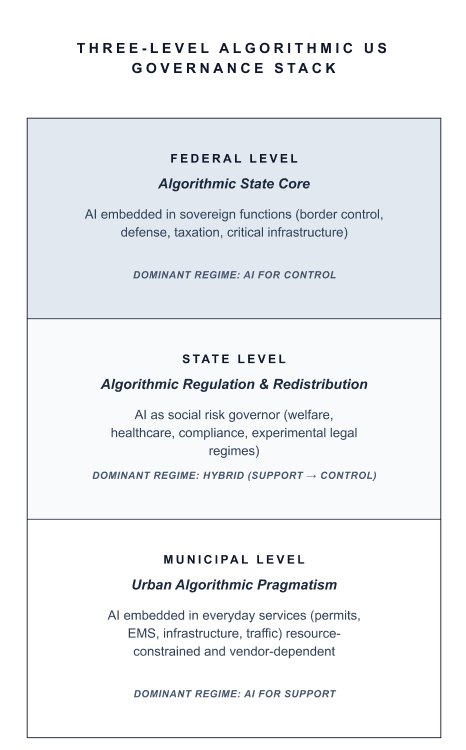}
  \caption{Three-Level Structure of Algorithmic Governance in the United States}
  \label{fig:three-level-stack}
\end{figure}

At the federal level, AI systems are consistently embedded in sovereign and high-stakes state functions, including border control, national security, taxation, defense, and large-scale scientific and infrastructural governance. Across the federal subsample (10 cases), AI deployment exhibits a high degree of institutionalization, operational scale, and political mandate.

Empirically, federal AI systems demonstrate the following recurrent characteristics:
\begin{itemize}
  \item High degree of institutional integration: AI systems function as persistent infrastructure rather than pilot tools.
  \item Default automation logics in selected domains, where algorithmic outputs serve as primary decision inputs.
  \item Strong reliance on centralized data architectures and commercial cloud vendors, resulting in hybrid public--private governance configurations.
  \item Dominant orientation toward control-oriented functions, including surveillance, risk detection, compliance enforcement, and strategic coordination.
\end{itemize}

Across the federal cases, AI is not limited to decision support but increasingly operates as a structural component of state capacity. The dominant regime at this level is empirically classified as \emph{AI for Control}, with supporting functions playing a secondary role.

At the state level (10 cases), AI systems primarily function as instruments for regulating access to social resources and managing social risk. The empirical material reveals a markedly different institutional role for AI compared to the federal level.

State-level AI deployments are characterized by:
\begin{itemize}
  \item Concentration in welfare, healthcare, labor, and social services, including Medicaid, unemployment insurance, homelessness prevention, and foster care systems.
  \item Hybrid governance functions, where AI simultaneously supports administrative decision-making and conditions access to public benefits.
  \item Greater legal and regulatory variability, including experimental legal regimes, regulatory sandboxes, and AI accountability statutes.
  \item Higher exposure to social fairness and distributional risks, particularly algorithmic bias and exclusion.
\end{itemize}

Empirically, state-level AI systems frequently mediate eligibility, prioritization, and intervention intensity, positioning algorithms as institutional gatekeepers rather than mere analytical tools. As shown in Figure~8, the dominant regime at this level is \emph{hybrid}, combining support-oriented analytics with control-oriented consequences for policy recipients.

At the municipal level (10 cases), AI adoption follows a distinct logic of operational pragmatism and service optimization. Municipal AI systems are embedded in everyday urban processes rather than strategic governance functions.

The empirical analysis identifies the following dominant features:
\begin{itemize}
  \item Frontline integration of AI into permitting, emergency response, infrastructure inspection, traffic management, and citizen communication.
  \item Predominance of modular, off-the-shelf AI solutions, with limited local customization.
  \item Strong vendor dependency, reflecting constrained technical and financial capacity.
  \item Short implementation cycles and experimental deployment practices.
  \item Minimal formal accountability infrastructure, including limited audit, appeal, and registry mechanisms.
\end{itemize}

In contrast to federal and state levels, municipal AI rarely automates binding decisions. Instead, algorithms reshape workflows, accelerate processing, and expand frontline capacity. Empirically, the dominant regime at this level is \emph{AI for Support}, with only a small subset of cases exhibiting control-oriented functions.

Across the full dataset, the study identifies a systematic functional differentiation of AI governance across levels of authority, summarized in Figure~8:
\begin{itemize}
  \item Control-oriented AI concentrates at the federal level.
  \item Redistributive and regulatory AI concentrates at the state level.
  \item Service-oriented and operational AI dominates at the municipal level.
\end{itemize}

Crucially, this differentiation is not driven by technological variation. Similar AI techniques---risk scoring, anomaly detection, predictive modeling, computer vision---are deployed across all three levels. The observed variation arises from differences in institutional mandate, accountability structures, resource availability, and normative expectations.

Beyond descriptive patterns, the study produces several analytical metrics that structure cross-case comparison:
\begin{enumerate}
  \item Degree of institutional embedding (from pilot tools to core governance infrastructure).
  \item Dominant governance function (control-oriented vs. support-oriented).
  \item Level-specific risk concentration, including sovereignty risks (federal), fairness risks (state), and accountability and vendor-dependency risks (municipal).
  \item Mode of algorithmic integration, ranging from advisory analytics to operational and infrastructural embedding.
\end{enumerate}
These metrics enable systematic comparison across heterogeneous cases without reducing AI governance to technical characteristics alone.

\section{Discussion}
This research aimed to clarify how artificial intelligence is institutionally embedded in a multi-level system of public authority and how this embedding leads to specific regimes of algorithmic governance. Through a comparative analysis of thirty cases of AI adoption in the United States at the federal, state, and municipal levels, this research shows that AI adoption in public governance does not follow a single path of ``algorithmic statehood.'' Rather, it follows level-specific paths of institutional embedding that systematically condition the functions, risks, and governance outcomes of algorithmic systems.

The main empirical result of this research is that there is a structurally differentiated governance stack, where the same types of AI systems play qualitatively different roles depending on their level of institutional embedding. Federal AI systems are mainly embedded in sovereign and high-risk tasks and are governed by control-oriented regimes. State-level AI systems are hybrid instruments that mediate access to social resources and govern social risk by combining support-oriented analytics with control-like distributive properties. Municipal AI systems are mainly support-oriented and aim to operationalize service delivery by transforming frontline administrative practices under conditions of limited institutionalization and fragmented accountability.

Most importantly, this differentiation cannot be accounted for by technology, policy type, or resource availability. Risk scoring, predictive analytics, computer vision, and generative models are used at all three levels. What differs is the institutional logic of their embedding, namely the mandate under which they operate, the accountability mechanisms that constrain them, and the normative expectations associated with their use. This finding supports the sociotechnical perspective on AI-enabled governance and specifies the causal role of institutional level as a structuring condition rather than a descriptive attribute.

Theoretically, the results specify the meaning of the ``third wave'' of digital governance by adding an intergovernmental component. The AI-enabled governance in the U.S. is no longer a homogenous change in the state but a set of parallel institutional developments across levels of government. Federal agencies are close to the maximum level of algorithmic autonomy, while states and local governments are still predominantly found in analytically supportive or hybrid regimes. The ``third wave'' of digital governance is no longer a homogeneous change but an uneven process, which is stratified by federalism.

The research also advances the sociotechnical approach to AI adoption by illustrating that algorithmic systems are less about discrete technological objects and more about institutional triggers. AI governance transforms the state not by substituting human decision-makers but by rebalancing data architectures, discretion, and power towards earlier stages of decision design and model governance. Differences in data centralization, vendor lock-in, and legal embeddedness account for why similar AI systems have different governance outcomes across levels.

The analytical distinction between control-oriented and support-oriented AI regimes is found to be an important interpretive finding. Empirical data reveals that this distinction does not map in a mechanical way to levels of government or types of technical systems. Rather, it identifies two normatively competing logics of algorithmic governance.

Control-oriented AI embodies a logic of discipline: algorithms are filters of suspicion, instruments of compliance, and devices for scaling surveillance and sanctioning power. Support-oriented AI, on the other hand, embodies a logic of solidarity: algorithms extend institutional visibility, facilitate preventive action, and reallocate administrative capacity to vulnerable groups. Crucially, the same technical infrastructure, such as predictive analytics or anomaly detection can embody either logic depending on design choices made within institutions.

In this way, AI governance appears as a normative choice realized through computation. The distribution of computational resources indexes underlying political values about control, redistribution, and tolerable state action. Algorithmic governance is thus not simply administrative or technical, but rather always already political in its effects.

Although the results provide a strong characterization of AI governance in the U.S. federal system, they are institutionally circumscribed. The theoretical model constructed in this research is unlikely to apply in full to:
\begin{itemize}
  \item Unitary states, in which the centralized state precludes cross-level differentiation and thus institutional variation in AI use.
  \item EU-style regulatory systems, in which supranational legal harmonization, ex-ante conformity assessment, and rights-based AI regulation preclude the development of multiple governance regimes.
  \item Low-capacity states, in which data infrastructure, administrative fragmentation, and sanctioning weakness preclude the development of either control-oriented or support-oriented algorithmic institutionalization.
\end{itemize}

In contrast, the results are most likely to spread to other federal or quasi-federal systems that are marked by decentralized power, institutional pluralism, and unequal administrative capacity, such as in Canada, Australia, India, or Brazil. In these cases, AI is also likely to solidify into differentiated regimes of governance that are determined by institutional level rather than technological design.

From a methodological perspective, the research shows that comparative multi-level qualitative case analysis can be a useful tool for analyzing algorithmic governance. By coding cases according to governance function, institutional embedding, and risk concentration, the research offers a framework for cross-national and cross-sectoral comparison that is replicable without reducing AI governance to diffusion rates or performance data.

From a practical perspective, the results of the research emphasize the importance of level-sensitive AI governance strategies. Federal governance structures that are suited to high-risk and centralized AI systems are not suitable for municipal governance, where vendor reliance and lack of accountability infrastructure are the norm. At the same time, service-oriented municipal AI systems pose their own risks, particularly with regard to transparency, appealability, and socio-spatial inequality, which are unlikely to be mitigated by federal standards.

The research also points to a number of areas that are ripe for further research. Quantitative research can complement the findings of the research by examining the prevalence and policy impact of different AI regimes. More detailed technical analyses of training data, calibration, and performance can help to shed light on how institutional design influences algorithmic behavior. Equally important, research on perception and interaction can help to examine how citizens and frontline officials experience algorithmic governance. Finally, systematic international comparisons can help to further illuminate the relationship between federalism, regulatory design, and state capacity in shaping the evolution of algorithmic governance.

Taken together, the results show that AI in U.S. public governance is more than a technological trend, but a contested institutional space shaped by federalism, normative choice, and sociotechnical configuration. The future of algorithmic governance cannot be predicted by technological capability alone. It will depend on the functions that states decide to algorithmize, the risks that they are willing to take, and the accountability mechanisms that they establish in response. In this sense, AI systems are mirrors of competing visions of justice, responsibility, and state power in the digital age.

\begin{thebibliography}{99}

\bibitem{AdaLovelaceAINowOGP2021}
Ada Lovelace Institute, AI Now Institute, \& Open Government Partnership. (2021).
\textit{Algorithmic accountability for the public sector}. Ada Lovelace Institute.

\bibitem{Bovens2007}
Bovens, M. (2007).
Analysing and assessing accountability: A conceptual framework.
\textit{European Law Journal}, 13(4), 447--468.
\url{https://doi.org/10.1111/j.1468-0386.2007.00378.x}

\bibitem{CriadoSandovalAlmazanGilGarcia2025}
Criado, J. I., Sandoval-Almazán, R., \& Gil-Garcia, J. R. (2025).
Artificial intelligence and public administration: Understanding actors, governance, and policy from micro, meso, and macro perspectives.
\textit{Public Policy and Administration}, 40(2), 234--251.
\url{https://doi.org/10.1177/09520767241272921}

\bibitem{DeloitteInsights2023}
Deloitte Insights. (2023).
\textit{AI-augmented government: Using artificial intelligence to transform public sector operations}.

\bibitem{DunleavyMargetts2023}
Dunleavy, P., \& Margetts, H. (2023).
Data science, artificial intelligence and the third wave of digital era governance.
\textit{Public Policy and Administration}.
\url{https://doi.org/10.1177/09520767231198737}

\bibitem{EngstromHoSharkeyCuellar2020}
Engstrom, D. F., Ho, D. E., Sharkey, C. M., \& Cuéllar, M.-F. (2020).
\textit{Government by algorithm: Artificial intelligence in federal administrative agencies}.
Administrative Conference of the United States.

\bibitem{IDC2024}
IDC. (2024).
\textit{Worldwide artificial intelligence spending guide}. International Data Corporation.

\bibitem{JanssenMellouliOjo2024}
Janssen, M., Mellouli, S., \& Ojo, A. (2024).
Introduction to the issue on artificial intelligence in the public sector: Risks and benefits of AI for governments.
\textit{Digital Government: Research and Practice}, 5(1), Article 1.
\url{https://doi.org/10.1145/3636550}

\bibitem{LiuLinChen2019}
Liu, H.-W., Lin, C.-F., \& Chen, Y.-J. (2019).
Beyond State v. Loomis: Artificial intelligence, government algorithmization and accountability.
\textit{International Journal of Law and Information Technology}, 27(2), 122--141.
\url{https://doi.org/10.1093/ijlit/eaz001}

\bibitem{NIST2023}
National Institute of Standards and Technology. (2023).
\textit{Artificial intelligence risk management framework (AI RMF 1.0)}.
U.S. Department of Commerce.
\url{https://www.nist.gov/itl/ai-risk-management-framework}

\bibitem{PeetersSchuilenburg2023}
Peeters, R., \& Schuilenburg, M. (2023).
Algorithmic governance: Technology, knowledge and power.
In W. Housley, A. Edwards, R. Beneito-Montagut, \& R. Fitzgerald (Eds.),
\textit{The SAGE handbook of digital society} (pp. 439--457). SAGE.
\url{https://doi.org/10.4135/9781529783193.n25}

\bibitem{TangiRodriguezMullerJanssen2025}
Tangi, L., Rodríguez Müller, P., \& Janssen, M. (2025).
AI-augmented government transformation: Organisational transformation and the sociotechnical implications of artificial intelligence in public administrations.
\textit{Government Information Quarterly}, 42(3), 102055.
\url{https://doi.org/10.1016/j.giq.2025.102055}

\bibitem{ZuiderwijkChenSalem2021}
Zuiderwijk, A., Chen, Y.-C., \& Salem, F. (2021).
Implications of the use of artificial intelligence in public governance: A systematic literature review and a research agenda.
\textit{Government Information Quarterly}, 38(3), 101577.
\url{https://doi.org/10.1016/j.giq.2021.101577}

\end{thebibliography}
\end{document}